\documentclass{aastex}

\usepackage{color}

\title{Time-Frequency Analysis of Superorbital Modulation of X-ray Binary SMC X-1 by Hilbert-Huang Transform}

\author{Chin-Ping Hu, Yi Chou}
\affil{Graduate Institute of Astronomy, National Central University, Jhongli 32001, Taiwan}
\email{Hu: m929011@astro.ncu.edu.tw, Chou: yichou@astro.ncu.edu.tw}
\author{Ming-Chya Wu\altaffilmark{1,2}}
\affil{Research Center for Adaptive Data Analysis, National Central University, Jhongli 32001, Taiwan}
\and
\author{Ting-Chang Yang, Yi-Hao Su}
\affil{Graduate Institute of Astronomy, National Central University, Jhongli 32001, Taiwan}

\altaffiltext{1}{Institute of Physics, Academia Sinica, Nankang, Taipei 11529, Taiwan}
\altaffiltext{2}{Department of Physics, National Central University, Jhongli 32001, Taiwan}

\begin{document}

\begin{abstract}
The high-mass X-ray binary (HMXB) SMC X-1 exhibits a superorbital modulation with a dramatically varying period ranging between $\sim40 \rm d$ and $\sim60\rm d$. This research studies the time-frequency properties of the superorbital modulation of SMC X-1 based on the observations made by the All-Sky Monitor (ASM) onboard the Rossi X-ray Timing Explorer (RXTE).We analyzed the entire ASM database collected since 1996. The Hilbert-Huang Transform (HHT), developed for non-stationary and nonlinear time series analysis, was adopted to derive the instantaneous superorbital frequency. The resultant Hilbert spectrum is consistent with the dynamic power spectrum while it shows more detailed information in both the time and frequency domains. The RXTE observations manifest that the superorbital modulation period was mostly betweenn $\sim50$ d and $\sim65$ d, whenas it changed to $\sim45 \rm d$ around MJD 50,800 and MJD 54,000. Our analysis further indicates that the instantaneous frequency changed in a time scale of hundreds of days between $\sim$MJD 51,500 and $\sim$MJD 53,500. Based on the instantaneous phase defined by HHT, we folded the ASM light curve to derive a superorbital profile, from which an asymmetric feature and a low state with barely any X-ray emissions (lasting for $\sim$0.3 cycles) were observed. We also calculated the correlation between the mean period and the amplitude of the superorbital modulation. The result is similar to the recently discovered relationship between the superorbital cycle length and the mean X-ray flux for Her X-1.
\end{abstract}


\section{Introduction}

Many X-ray binaries exhibit long-term modulation with periods that are longer than their orbital periods. The modulation time scales of superorbital periods are typically tens to hundreds of days. A few X-ray binaries such as Her X-1 \citep{Tananbaum1972} show superorbital modulation with stable periodicities, though recent discoveries show that the superorbital cycle length may vary within a small range \citep{Leahy2010}. A small group of X-ray binaries show quasi-periodic superorbital modulation with periods changing within a certain range. The most prominent of these is SMC X-1 \citep{Clarkson2003}. Some other X-ray binaries such as Cyg X-2 exhibit superorbital modulation with multi-periodic modulations  \citep{Clarkson2003b}. Such long-term superorbital variations may be caused by the precession of a tilted or warped accretion disk \citep{Wijers1999, Ogilvie2001}, a hierarchical third companion like 4U 1820-30 \citep{Chou2001}, or by reasons currently unknown.

The high-mass X-ray binary (HMXB) SMC X-1, which was first discovered in 1971 \citep{Leong1971}, consists of a 1.06 $\rm M_{\odot}$ neutron star \citep{vandermeer2007} and a B0 I type supergiant with a mass of 17.2 $\rm M_{\odot}$ \citep{Reynolds1993}. The spin period of the neutron star is 0.71 s and it decreases with time \citep{Wojdowski1998}. The orbital period of this system is $\sim 3.89$ d as evaluated by its eclipse \citep{Schreier1972}, and the orbital inclination angle is $\sim67^{\circ}$ \citep{vandermeer2007}. By analyzing the mid-eclipse time delay, \citet{Wojdowski1998} found that the orbital period also changes with time ($\dot{P}_{orb}/P_{orb}\approx -3.35\times10^{-6} \rm yr^{-1}$).

From the observational data collected by the High Energy Astronomy Observatory 1 (HEAO-1), \citet{Gruber1984} found that the X-ray flux of SMC X-1 varies with a rough time scale of 60 d. \citet{Wojdowski1998} confirmed this periodicity and showed that the superorbital cycle length was not stable during the first $\sim 500$ d observations made by the All-Sky Monitor (ASM) onboard the Rossi X-ray Timing Explorer (RXTE). After the ASM collected sufficient amounts of data, several time-frequency analysis methods such as the wavelet transform \citep{Ribo2001}, dynamic power spectrum \citep{Clarkson2003}, and sliding Lomb-Scargle periodogram \citep{Trowbridge2007} were applied to the light curve to investigate the variations in the superorbital period of SMC X-1. The mechanism of superorbital modulations in SMC X-1, similar to those in Her X-1, LMC X-4, and Cyg X-2, are interpreted by a warped and tilted accretion disk \citep{Wojdowski1998}. When the disk precesses, it obscures our line of view to the central X-ray source. However, the mechanisms that cause the precession period to change with time are still unknown.

\citet{Ribo2001} analyzed the first $\sim 1700$ d of the ASM light curve and found that the superorbital period first decreases from $\sim60$ d to $\sim45$ d and subsequently increases to $\sim60$ d. By fitting the maximum points of the wavelet spectrum with a sinusoidal curve, they found that the modulation period varies with a period of $1421\pm8$ d. \citet{Clarkson2003} analyzed the first $\sim2200$ d of the ASM light curve along with all the Burst and Transient Source Experiment (BATSE) data using the dynamic power spectral technique. The dynamic power spectra of both the RXTE and BATSE light curves show that the superorbital modulation period changes between $\sim 60$ d and $\sim 45$ d. The superorbital period appears to vary in a $\sim 1600$ d time scale. \citet{Trowbridge2007} analyzed the first $\sim$4000 d of ASM data using the sliding Lomb-Scargle periodogram and the the cycle length counting method, and the results indicated that the cycle length variation is perhaps not as smooth as that obtained from the dynamic power spectrum. The cycle length variation possibly contains a shorter periodicity in addition to the $\sim1600$ d period. However, due to the constraints of the window size in both the dynamic power spectrum and the sliding Lomb-Scargle periodogram, it is difficult to further analyze the periodicities of the superorbital period variation.

In order to study the superorbital period variation with sufficient statistical accuracy, here we adopt the Hilbert-Huang transform (HHT) \citep{Huang1998} to derive instantaneous frequency of the ASM light curve data for periodicity analysis. HHT is a new time-frequency analysis method, designed for non-stationary and nonlinear time series analysis. It consists of the empirical mode decomposition (EMD) and the Hilbert spectral analysis. In this work, we first apply EMD to decompose the ASM light curve into a number of intrinsic mode functions (IMFs) which are suitable for the Hilbert transform, and then calculate the instantaneous frequency and amplitude of certain IMFs, which form the basis of further analysis on periodicity of superorbital period variation, folded light curve, and correlations between the period and the amplitude. The proposed analysis allows us to explore subtle variations in the superorbital periods which were not detected by existing methods.

The rest of this paper is organized as follows. In section \ref{obs}, we briefly introduce the ASM observations and the algorithm of HHT. Section \ref{result} describes the time-frequency analysis of ASM observations based on HHT. Finally, we discuss our results in section \ref{discussion}.

\section{Observation and Data Analysis}\label{obs}

\subsection{RXTE ASM}
Since the RXTE was launched in late 1995, the ASM continuously sweeps the entire sky once every 90 minutes. The energy range of the ASM is 1.3 to 12.1 keV. This can be further divided into three energy channels (ch1: 1.3-3.0 keV, ch2: 3.0-5.0 keV, and ch3: 5.0-12.1 keV). The summed band data collected from 1996 to 2010 (MJD 50134 to MJD 55371), with a total time span of $\sim5000$ d were used in this time-frequency analysis. The dwell data, where all the eclipses were removed according to the ephemeris proposed by \citet{Wojdowski1998}, were binned into a one-day averaged light curve. Following the method of \citet{Trowbridge2007}, we did not exclude those data points that have a count rate falling below zero. Figure \ref{lightcurve} shows the ASM light curve of SMC X-1. Although the dwell data were binned into a one-day averaged light curve, it still contained some missing points or small observation gaps. We then used the piecewise cubic Hermite interpolation \citep{Kahaner1989} to interpolate the ASM light curve to evenly sampled data.

\subsection{Hilbert-Huang Transform}
Owing to the high variability of the superorbital cycle length of SMC X-1, it is suitable to analyze the time-frequency properties of its superorbital modulation by HHT. We use the Hilbert transform to calculate instantaneous frequency of the data. For a time series $x(t)$, this is achieved by first calculating its conjugate $y(t)$, defined as
\begin{equation}
y(t)=\frac{1}{\pi}P\int_{-\infty}^{\infty} \frac{x(t)}{t-t'}dt',
\end{equation}
where $P$ indicates the Cauchy principal value. In the complex plane, we define an analytic signal, $z(t)$, as
\begin{equation}
z(t)=x(t)+iy(t)=a(t)e^{i\theta(t)},
\end{equation}
where $a(t)=\sqrt{x^{2}(t)+y^{2}(t)}$ is the amplitude and $\theta(t)$ is the angular phase function. The instantaneous angular frequency $\omega(t)$ is then defined as the time derivative of $\theta(t)$,
\begin{equation}
\omega(t)=\frac{d\theta(t)}{dt}.
\end{equation}
Accordingly, we can obtain the frequency instantaneously as long as the sampling rate is sufficiently high. However, the instantaneous frequency is meaningful only if the time series $x(t)$ is an IMF \citep{Huang1998}. An IMF satisfies the following two conditions: (1) the number of extrema and the number of zero crossings must be identical or differ by one; and (2) at any of the points in the series, the mean value of the upper envelope (defined by the local maxima) and the lower envelope (defined by the local minima) is zero \citep{Huang1998}. Because the real data are usually not IMF, it is difficult to directly apply the Hilbert transform on primary data in most cases. To address this issue, \citet{Huang1998} proposed the EMD method to decompose a time series into several IMFs such that the Hilbert Transform is applicable.

The EMD method assumes that any time series consists of some oscillatory components. The decomposition scheme utilizes the actual time series for the construction of the decomposition base rather than decomposing it into a prescribed set of base functions. The decomposition is achieved by iterative ``sifting'' processes for extracting modes by identification of local extrema and subtraction of local means \citep{Huang1998}. The iterations are terminated by a criterion of convergence. For details of the sifting, reference is here made to \citet{Huang1998}. Under the procedures of EMD, the original time series $x(t)$ is decomposed according to
\begin{equation}
x(t)=\sum_{i=1}^{n}c_{i}+r_{n},
\end{equation}
where $c_i$'s are IMFs and $r_n$ is a residue. For a time series with infinite length, the decomposition satisfies the properties that the decomposed components are orthogonal to each other and form a complete set. For a time series with finite length; however, the orthogonality of the decomposition may not strictly hold. In this case, we minimize the orthogonality index (OI) defined by
\begin{equation}
\mathrm{OI} = \frac{\sum_t\sum_{i\neq j} c_i(t)c_j(t)}{\sum_t x^2(t)},
\end{equation}
to make the decomposition ``unique''. Furthermore, if the original data is noisy or contains high-frequency intermittent signals, the decomposition may suffer the mode mixing problem, \textit{i.e.} the modulation signal with the same time scale across different IMFs \citep{Huang1998}. This would cause many spurious, confused signals in the transition region. When the mode mixing problem occurs, it is difficult to obtain a rigorous instantaneous frequency, $\omega(t)$, of the modulation to be further analyzed. Then, the ensemble EMD (EEMD) \citep{Wu2009}, a revised version of EMD implemented by taking an ensemble average of the components decomposed from combinations of primary time series and distinct white noises, is used to mitigate the mode mixing problem, and a post-processing EMD \citep{Wu2009} is applied afterward on the outcome of EEMD to guarantee that the final IMFs fulfil orthogonality and completeness.

After decomposing the time series into IMFs, the Hilbert transform is applied on each IMF. As a result, the data can be expressed as
\begin{equation}
x(t)=\sum_{j=1}^n a_j(t)\exp\left( {i\int ^t _0 \omega_j(t^{\prime})dt^{\prime}}\right),
\label{hht}
\end{equation}
where $a_j(t)$ and $\omega_j(t)$ are the instantaneous amplitudes and frequencies of decomposed components. The amplitude and frequency of each component in Eq.(\ref{hht}) are functions of time, instead of constant amplitude and constant frequency in the Fourier transform.

Figure \ref{eemd} shows the decomposition of the interpolated ASM light curve of SMC X-1. There are $11$ IMFs, $1$ residue component, and $\mathrm{OI}=0.036$. The significance test \citep{Wu2004} on these IMFs suggests that the 40--60 d superorbital modulation signal of SMC X-1 mostly concentrates in the fifth component ($c_5$). The values of all the other components are lower than or marginal at the $3\sigma$ white noise level, suggesting that they act in a manner similar to white noise. These components are probably caused by the observational noise associated with non-periodic modulations in other time scales.

\section{Results}\label{result}
\subsection{Hilbert Spectrum}
After decomposing the ASM light curve into IMFs, the normalized Hilbert transform \citep{Huang2003} was applied on the IMFs to obtain the instantaneous frequencies and amplitudes. The normalized Hilbert transform was proposed to overcome the limitation of absence of strong amplitude modulations imposed by the Bedrosian theorem {\citep{Bedrosian1963}. The resultant Hilbert spectrum is a three-dimensional map which displays how the modulation period and amplitude vary with time. Figure \ref{hilbert_dynamic} shows the result, in which the frequency range is divided into 3000 bins and the spectrum is smoothed by a Gaussian filter for clarity. The color map represents the Hilbert energy spectrum with the magnitude of energy defined as square of the amplitude. For comparison, the dynamic power spectral technique, as described in \citet{Clarkson2003}, was also applied on the ASM light curve. We first calculated the Lomb-Scargle power spectrum \citep{Scargle1982} of the first 200 d of the ASM data, then successively moved the 200 d data window forward by a step size of 10 d and applied the Lomb-Scargle algorithm on each segment to obtain a series of power spectra. The power spectra are plotted in Figure \ref{hilbert_dynamic} in the contour form with the spectral power as defined in the Lomb-Scargle periodogram \citep{Scargle1982}. It is obvious that the Hilbert energy spectrum is consistent with the dynamic power spectrum. Because the spectral analysis based on HHT is independent of window size, more detailed structures in the light curve data can be observed from the high-resolution Hilbert spectrum than the dynamic power spectrum.

The first remarkable feature of both spectra is that the superorbital modulation period changes dramatically from $\sim 60$ d to $\sim 45$ d and goes back to $\sim 60$ d between $\sim$MJD 50,200 and $\sim$MJD 51,100. This feature is consistent with previous studies \citep{Ribo2001, Clarkson2003, Trowbridge2007}. Interestingly, a similar variation in the superorbital modulation period is repeated between $\sim$MJD 53,600 and $\sim$MJD 54,400. The time durations of these two events are roughly 800 d and the separation between them is $\sim 3200$ d. This indicates that such a phenomenon is probably recurrent. In addition, although the superorbital period is relatively stable, it seems to oscillate between $P\sim 50$ d and $\sim65$ d over a time scale of hundreds of days between $\sim$MJD 52,000 and $\sim$MJD 53,000. This feature can be seen very marginally in the dynamic power spectrum, but it is difficult to identify due to the limitation of the window size. By contrast, the Hilbert spectrum based on HHT displays the instantaneous frequency (or period) more accurately, enabling the application of further timing analysis methods on the instantaneous frequency to assess the periodicity hidden in the superorbital period change.

In order to study the periodicities in the variation of the superorbital frequency, the Lomb-Scargle periodogram was applied on the instantaneous frequency of the component $c_5$, and the result is shown in Figure \ref{if_ls_period}. The highest peak is located at $P=1674$ d; this is consistent with the previously reported periodicity \citep{Ribo2001,Clarkson2003}. However, this peak is the second harmonic of $P\sim3200$ d, and it corresponds to the separation of the events occurring when the superorbital period changes between $\sim 60$ d and $\sim 45$ d. Moreover, a few peaks with relatively less power values located at $P\sim240$ d and $P\sim320$ d are observed. These peaks are likely due to the oscillation feature mentioned in the previous paragraph.

It is possible to further apply time-frequency analysis methods such as HHT and the dynamic power spectrum on the instantaneous frequency to examine the variation of the superorbital modulation with time. Figure \ref{if_dynamic} shows a dynamic power spectrum with a window size of 1,000 d and a movement of 10 d. The most prominent peaks are located at around MJD 50,800 and MJD 54,000. These peaks correspond to the short-period state of the superorbital modulation in the time domain in Figure \ref{hilbert_dynamic} and the highest peaks in the frequency domain in Figure \ref{if_ls_period}. During $\sim$ MJD 52,000 and $\sim$ MJD 53,500, the most prominent periodicities disappear and smaller periodicities from $P\sim240$ d to $P\sim320$ d are observed. These correspond to the oscillation feature of the instantaneous frequency with a time scale of hundreds of days, as shown in Figure \ref{hilbert_dynamic}. Figure \ref{if_dynamic} shows that the oscillation feature seems to appear again after $\sim$ MJD 54,500, indicating that it might exist in the entire observation time span, except during the two events where the period changes dramatically between $\sim 60$ d and $\sim 45$ d.

\subsection{Superorbital Modulation Profile}
It is difficult to derive a proper profile to describe the characteristics of superorbital modulation for highly variable cycle lengths from a light curve folded with a fixed period. \citet{Clarkson2003} first selected a segment of light curve at which the frequency seems stable in the dynamic power spectrum and subsequently folded it with a fixed period. However, the period still changes during the entire observation time even in the stable region of the dynamic power spectrum, according to the Hilbert spectrum and the analysis of cycle length counting by \citet{Trowbridge2007}. As an improvement, \citet{Trowbridge2007} fitted the smoothed light curve with fourth-order polynomial functions to obtain the minimum points of the superorbital modulation, and folded the whole light curve according to the ephemeris defined by minima. The difficulty then becomes the identification of some minima for the presence of noise in the ASM light curve and distortions of the modulation shape when the period changes dramatically. By contrast, the light curve folded according to the phase defined as $\phi(t)=\theta(t)/2\pi$ in HHT does not suffer the above-mentioned problems. Figure \ref{fold_lc} shows the folded light curve of SMC X-1, and the phase zero epoch is defined by the first data point in the light curve (MJD 50,134).

The folded light curve shown in Figure \ref{fold_lc} seems mostly in agreement with the results of \citet{Trowbridge2007}. The major distinction between this superorbital modulation profile and that obtained from previous studies is a clear low state with a negligible X-ray flux lasting for $\sim 0.3$ cycle. Moreover, the asymmetric features of the superorbital profile can also be obtained in the folded light curve. In order to describe the asymmetric profile, the low state count rate was defined as the mean count rate during phase $\sim 0.2$ to $\sim 0.55$ and the high state count rate was considered by averaging the count rate between phase  $\sim 0.75$ to $\sim 1.0$. Thus, the rising and falling time scale can be estimated by calculating the time scale between 10\% and 90\% of amplitude. By this definition, the rising time scale is $\sim0.14$ cycle (from phase 0.05 to 0.19), which is shorter than the falling time scale of $\sim0.19$ cycle (from phase 0.54 to 0.73).

The ASM light curve can further be divided into three channels. The spectral hardness provides crude information about the emission mechanisms during superorbital modulation. The ASM hardness ratio was defined as $ch3/(ch1+ch2)$, since $ch3$ and $ch1+ch2$ have similar photon count rates during the high state. Only the data with a signal to noise ratio (SNR) greater than 5 were chosen and folded according to the superorbital phase defined in HHT. During the entire observation time span, there are no data points with SNRs greater than 5 between phase 0.73 and 1.05. Therefore, no reliable hardness ratio in the low state was obtained. The folded hardness ratio is shown in Figure \ref{fold_hr}. It is easily observed that the hardness ratios in the high state (from phase 0.19 to 0.54) are relatively stable. In the transition state (phase 0.05--0.19, and 0.54--0.73), the hardness ratios do not appear to significantly deviate from the mean value as compared with that in the high state although the errors are much greater.

\subsection{Correlation between Period and Amplitude}
Another remarkable feature observed in this study is the correlation of the amplitude with the period. Using the instantaneous frequency and amplitude derived by HHT, the correlation between the cycle-averaged period and the Hilbert amplitude in the most significant IMF ($c_5$) was calculated. The rank correlation coefficient is $r=0.46$ with a null hypothesis probability value of $3.2\times10^{-6}$, indicating a significant correlation between the period and the amplitude (see Figure \ref{corr_mp_amp}). 

In order to obtain the correlation between the period and amplitude in the original data set, the rms amplitude of each cycle was calucalted. The result shows that the period and the rms amplitude still show a significant correlation ($r=0.32$ with null hypothesis probability value of $1.4\times 10^{-3}$, see Figure \ref{corr_mp_rmsamp}). Although the correlation appears to change to an anti-correlation after $P\gtrsim 60$ d, the number of cycles is too few to obtain a significant correlation for periods longer than $\sim60$ d. Similar correlation is observed in the superorbital modulation of Her X-1 \citep{Leahy2010, Still2004}.

\section{Discussion}\label{discussion}

We have successfully performed HHT-based time-frequency analysis on the ASM light curve of SMC X-1. The Hilbert spectrum manifests variations in the instantaneous period of the superorbital modulation. The instantaneous phase defined by HHT allows us to fold the light curve to derive a reasonable superorbital profile. The time-frequency information obtained from the HHT analysis is more abundant than the traditional Fourier-based methods. This benefits our investigations on the superorbital modulation of SMC X-1.

Recent observations by the Chandra and XMM-Newton show that the emission of SMC X-1 could be described as a blackbody component with $kT_{BB}\sim0.18$ keV and a hard power law component \citep{Neilsen2004}. The emissions with energy $\lesssim 1.0$ keV are dominated by the blackbody radiation. These emissions are reprocessed X-rays emerging from the illuminated warp region of the accretion disk \citep{Hickox2005}. On the other hand, the high energy emissions are dominated by the power law components owing to pulsar beams emerging from the neutron star surface. Therefore, the energy range of all the ASM channels is dominated by the power law components. From the folded hardness ratio, we could not obtain the absorption feature since the hardness ratios in the transition state do not significantly differ from that in the high state. This indicates that the warp region of the accretion disk is optically thick for all the ASM energy bands, and the superorbital profile is determined by the covering of the central source by the warp region.

Furthermore, we found a significant correlation between the amplitude and the period of the superorbital modulation. A similar correlation is observed in Her X-1, which is a Low-Mass X-ray Binary (LMXB) with an orbital period of 1.7 d and shows a relatively stable superorbital modulation in a 35 d period \citep{Tananbaum1972}. From the ephemeris obtained by the O$-$C method, \citet{Still2004} showed that the time-averaged peak main-on flux and the precession period have a positive correlation though the quantity of obtained data is very small. \citet{Leahy2010} analyzed the ASM light curve using the cross-correlation method and showed that the cycle length and average ASM count rate indeed have a significant correlation. Although the mechanisms that cause the disk precession period change with time are unknown, \citet{Wijers1999} showed that for a radiation-induced warp disk, the precession period of the outside edge of the disk would anti-correlate with the strength of the radiation field. From the observation that the low state of SMC X-1 has barely X-ray flux, the modulation amplitude might be associated with the central X-ray flux if the central emission source is totally uncovered during the high state. Our results for SMC X-1, as well as the results for Her X-1 \citep{Still2004, Leahy2010}, seem to be inconsistent with this prediction. \citet{Leahy2010} accounted for the average ASM count rate, superorbital cycle length, and spin-down of the neutron star in their studies and obtained a consistent result, indicating that the cycle length could directly be related with the variations in accretion flow. For SMC X-1, the evolution of the pulse frequency was demonstrated by \citet{Inam2010}. The frequency residual during the RXTE observations seems to vary with the time scale of the superorbital variation. However, the observed sampling is insufficient to obtain a significant correlation between the spin frequency residual and the superorbital period variation.

An alternative possibility is that the amplitude might be affected not only by the X-ray flux of the central source but also by the variations in tilt angle, warp mode, etc. \citet{Ogilvie2001} studied the stability of radiation-driven warping of accretion disks in X-ray binaries and found that the binary separation ($r_b$) and mass ratio ($q$) are good indicators of warping behavior. SMC X-1 lies in the marginal region of the $r_b-q$ parameter space where mode 0 stable warping and other warping modes can possibly appear \citep{Clarkson2003b}. In contrast to SMC X-1, Her X-1 and LMC X-4 exhibit only stable mode 0 warping. However, the ÔstableÕ warping in Her X-1 also shows cycle length variations. This indicates that mode 0 is perhaps not stable, and the correlation between cycle length and amplitude or mean count rate is a property of mode 0.

\acknowledgments
This research has made use of data provided by the ASM/RXTE teams at MIT and at the RXTE SOF and GOF at NASA's GSFC. This research is supported by grant NSC 97-2112-M-008-012-MY3 of the National Science Council of Taiwan.

\begin{figure}
\plotone{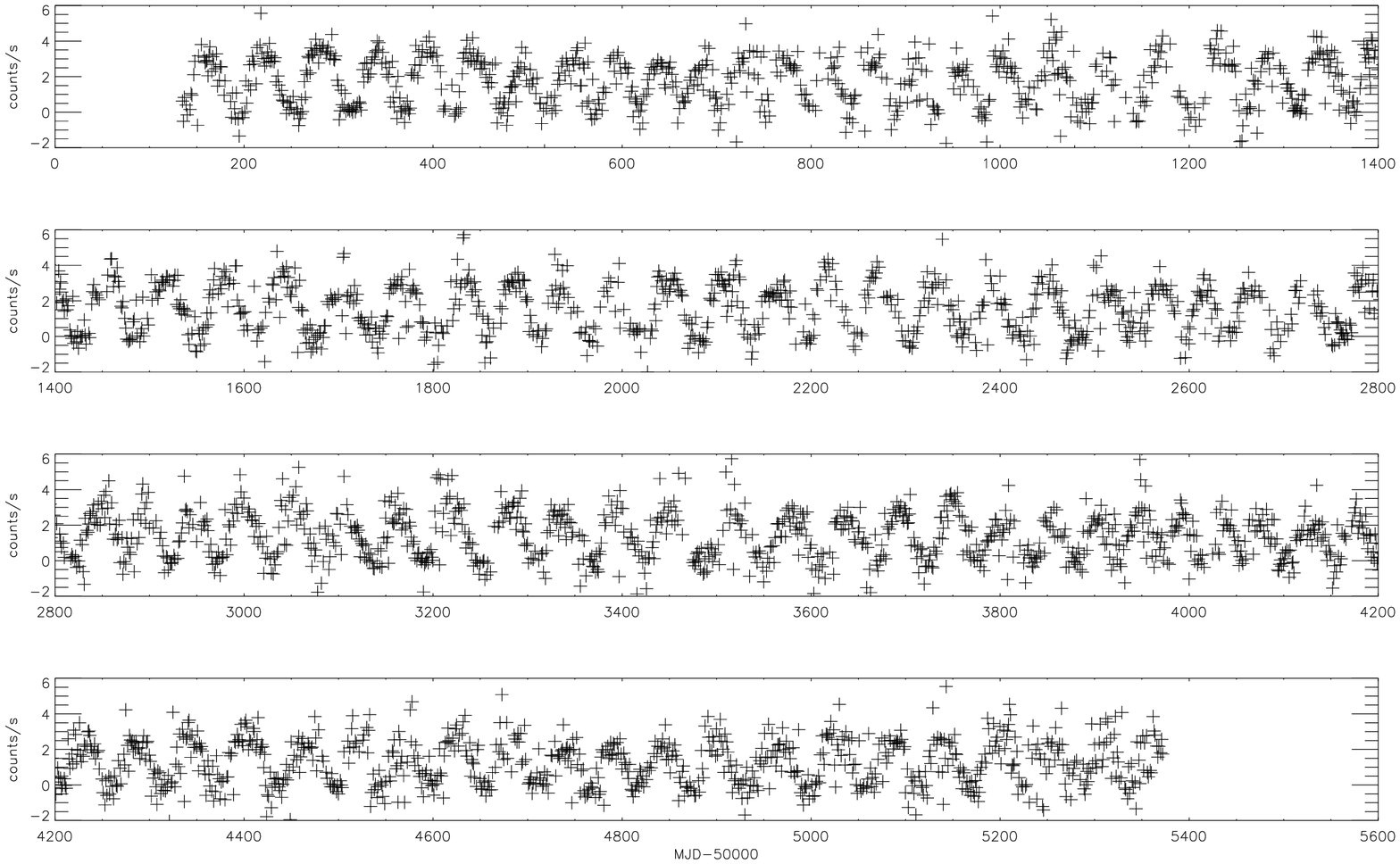} \caption{All-Sky Monitor (ASM) light curve of SMC X-1. } \label{lightcurve}
\end{figure}

\begin{figure}
\begin{center}
\includegraphics{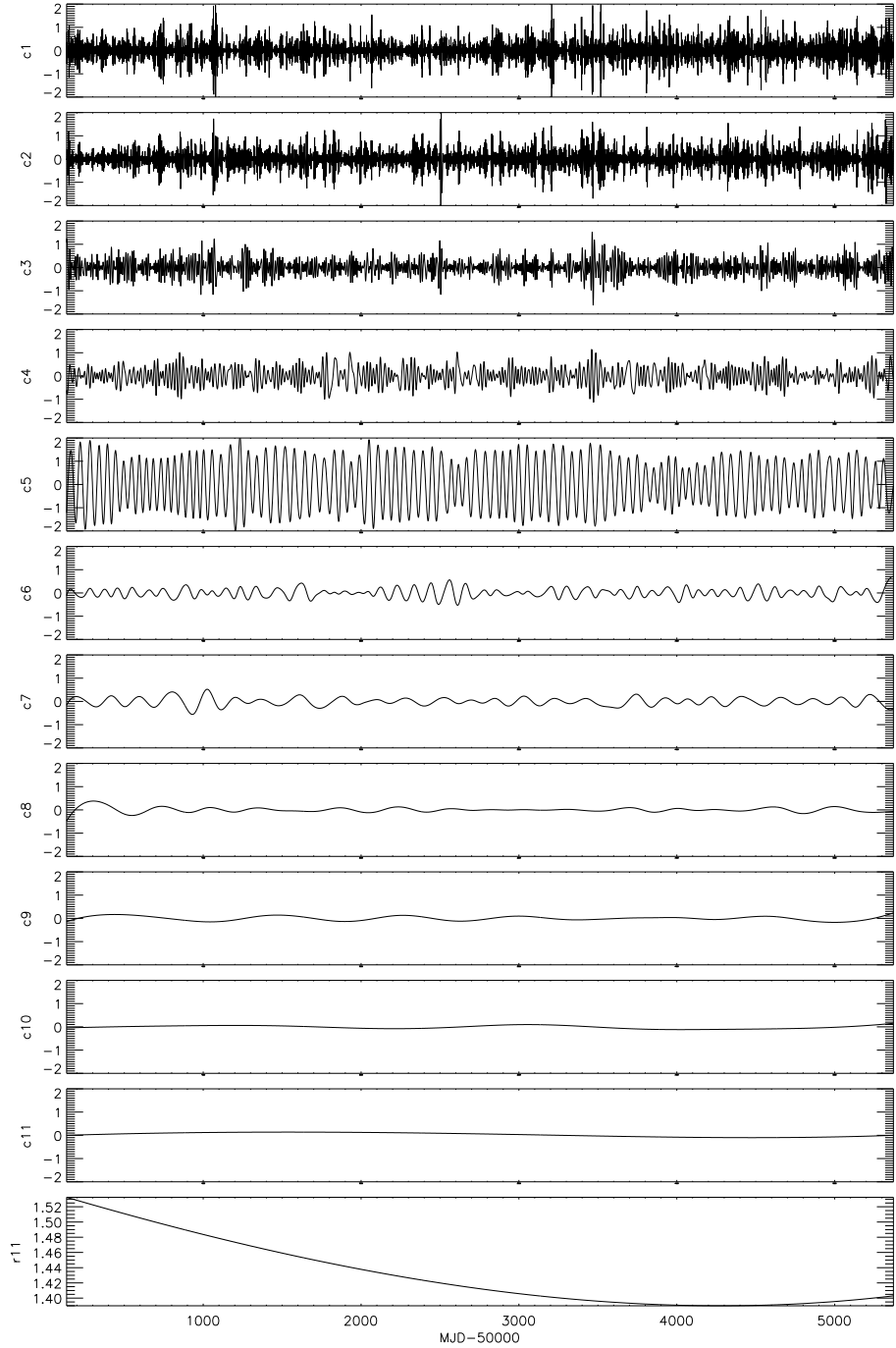} \caption{Intrinsic mode functions (IMFs) decomposed by ensemble empirical mode decomposition (EEMD). The fifth IMF ($c_5$) is responsible for the $\sim40$ d to $\sim60$ d superorbital modulation.} \label{eemd}
\end{center}
\end{figure}

\begin{figure}
\epsscale{1.0}
\plotone{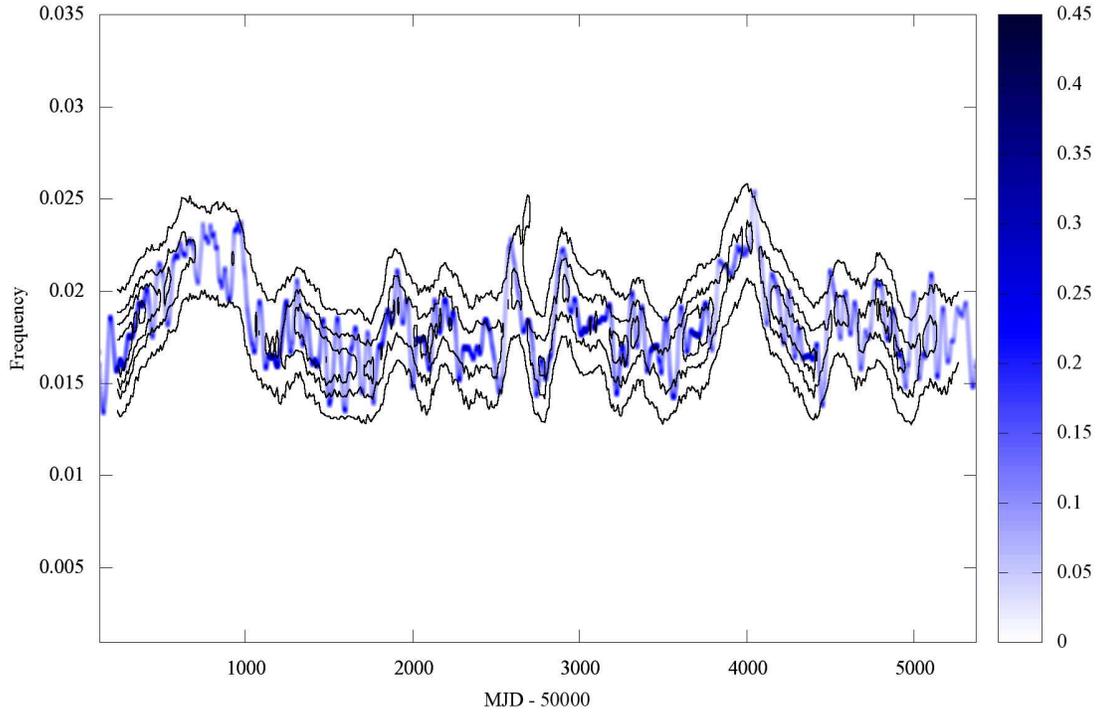} \caption{Hilbert energy spectrum (color map) and dynamic power spectrum (contour plot). The blue curve represents the instantaneous frequency of IMF  $c_5$ after Gaussian smoothing and the color depth denotes the magnitude of the Hilbert energy. The window size of dynamic power spectrum is 200 d and the movement is 10 d.} \label{hilbert_dynamic}
\end{figure}

\begin{figure}
\plotone{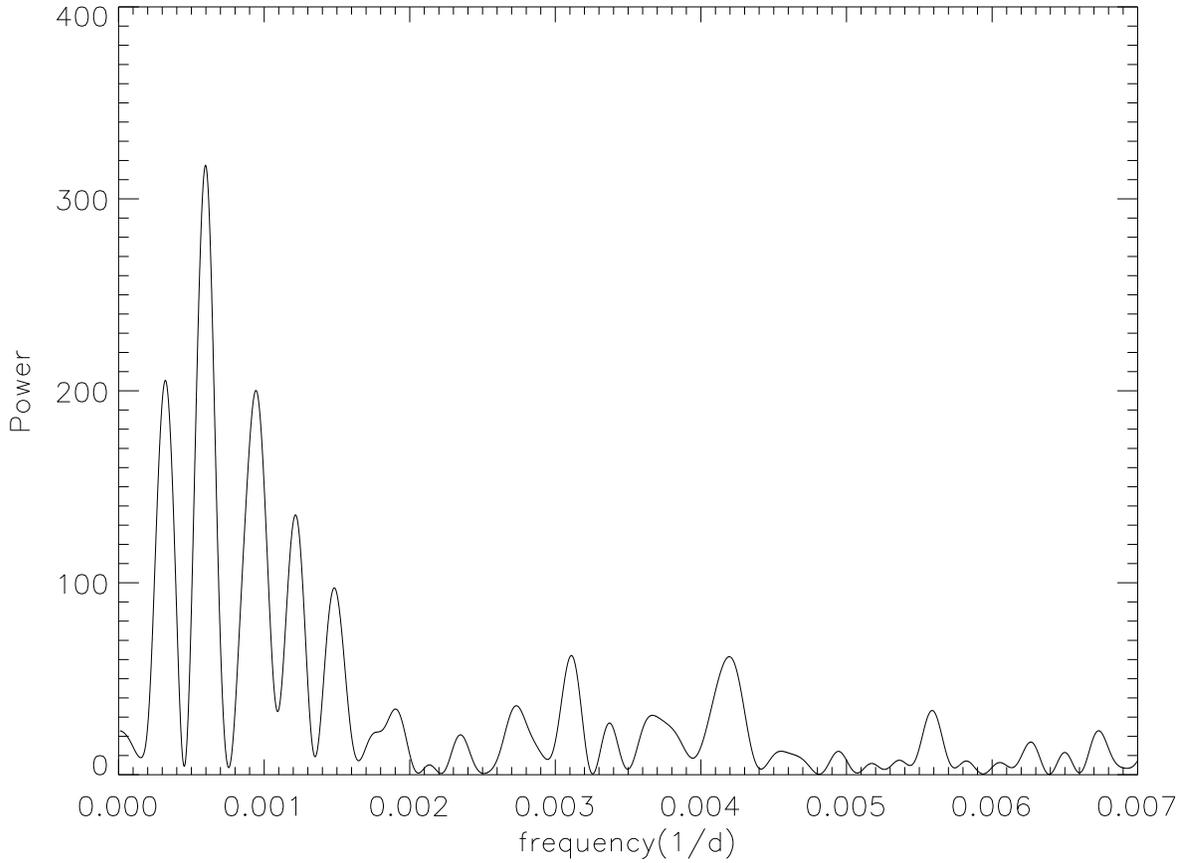} \caption{Lomb-Scargle periodogram of the instantaneous frequency of IMF $c_5$. The highest peak is located at $f\sim0.0006$ (or $P=1647$ d), which is the second harmonic of $P\sim 3200$ d peak. The other prominent peaks are located at $f\sim0.0031$ ($P\sim240$ d) and $f \sim0.0031$ ($P\sim320$ d).
\label{if_ls_period}}
\end{figure}

\begin{figure}
\plotone{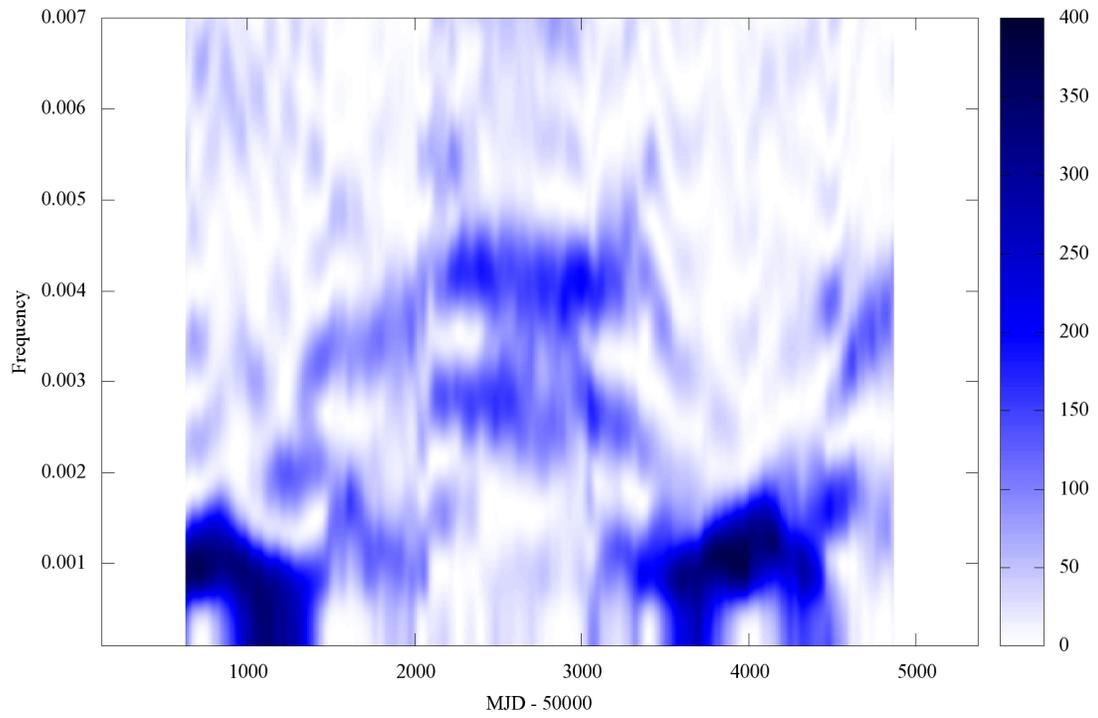}
\caption{Dynamic power spectrum of the instantaneous frequency of IMF $c_5$. \label{if_dynamic}}
\end{figure}

\begin{figure}
\plotone{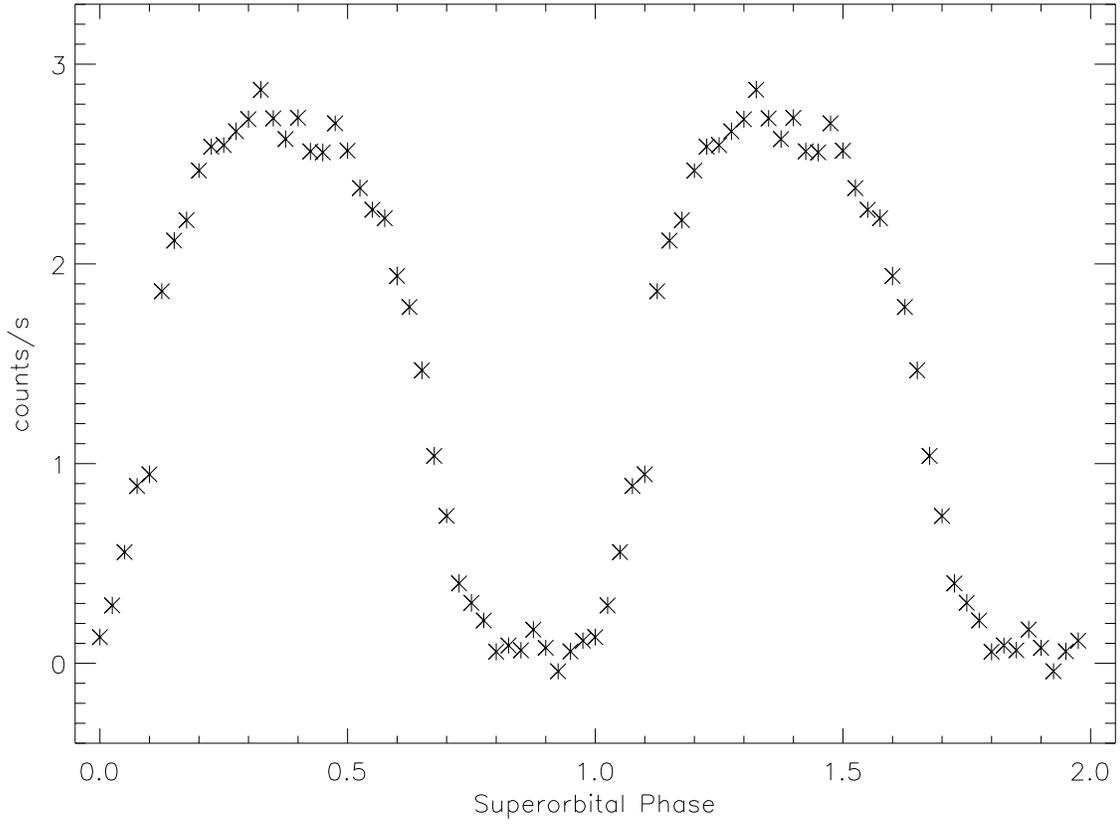}
\caption{Folded light curve of SMC X-1 according to the phase defined by HHT. The number of bins in one cycle is 40. \label{fold_lc}}
\end{figure}

\begin{figure}
\plotone{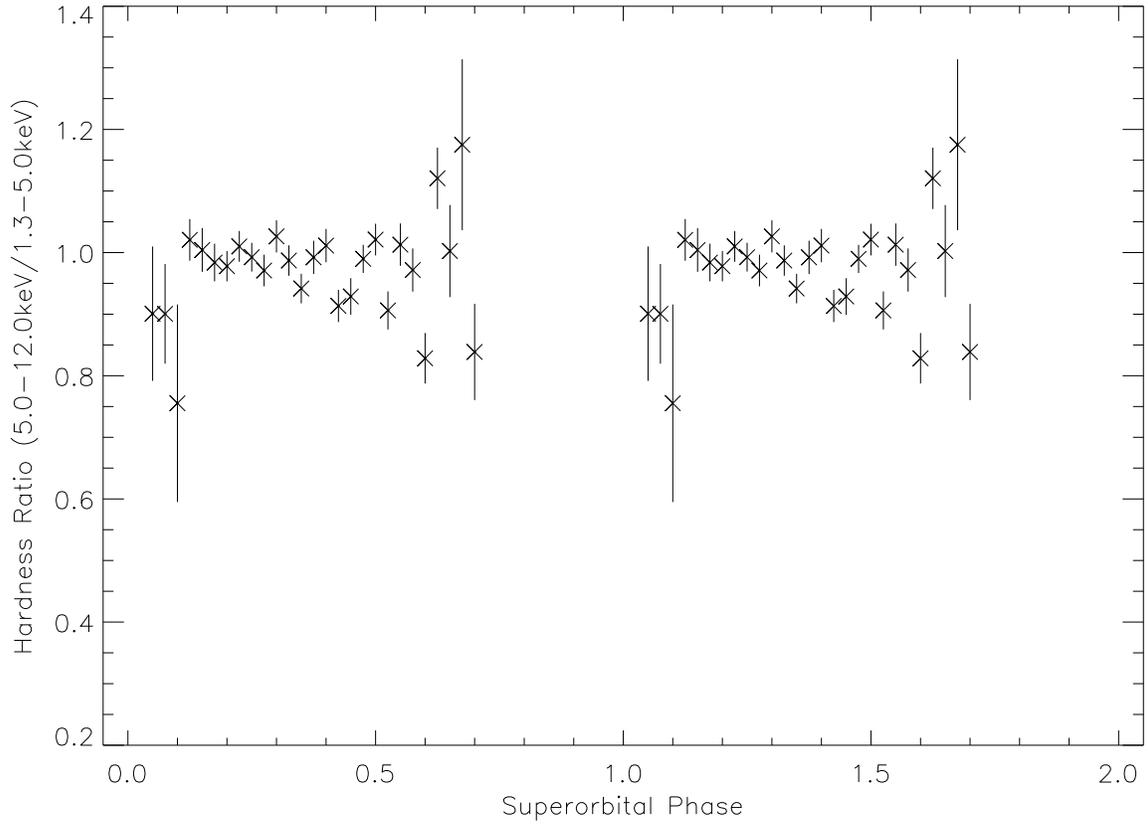}
\caption{Hardness ratios folded according to the superorbital phase. Only the data points with a signal to noise ratio greater than 5 were chosen to calculate the hardness ratios. \label{fold_hr}}
\end{figure}

\begin{figure}
\plotone{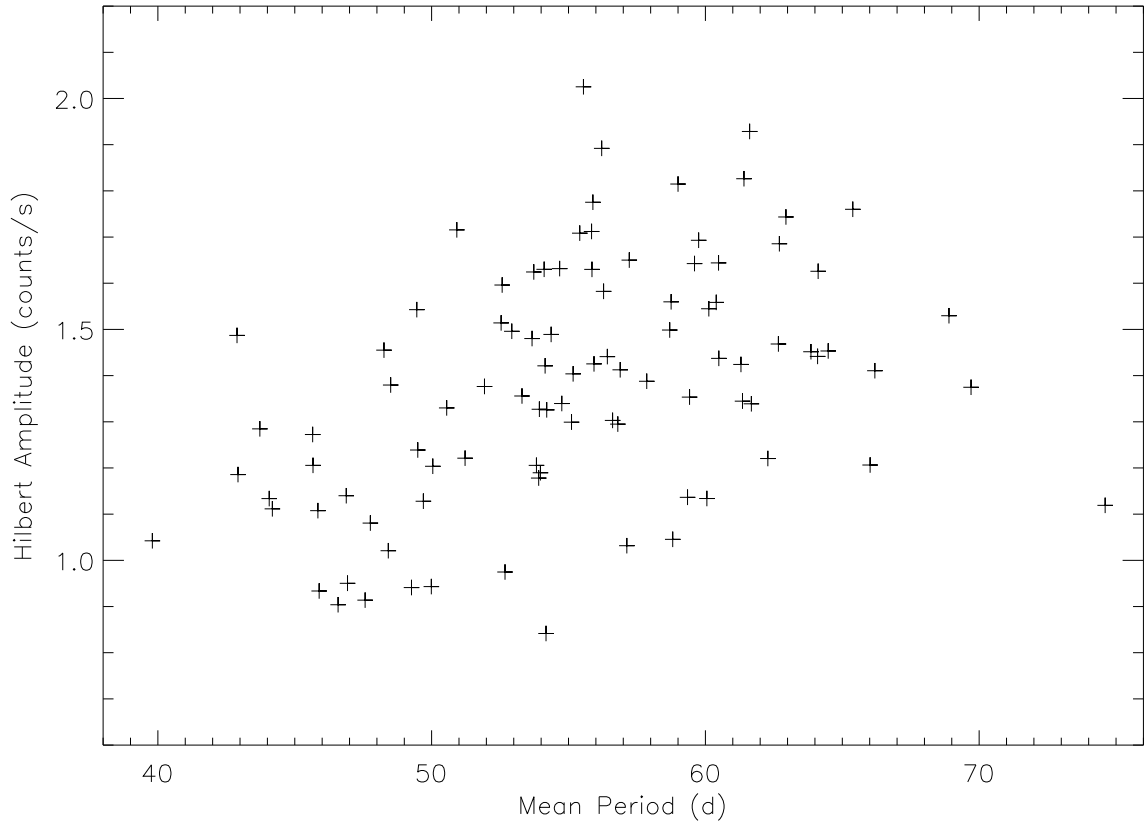}
\caption{Correlation of cycle-averaged period and Hilbert amplitude of IMF $c_5$. \label{corr_mp_amp}}
\end{figure}

\begin{figure}
\plotone{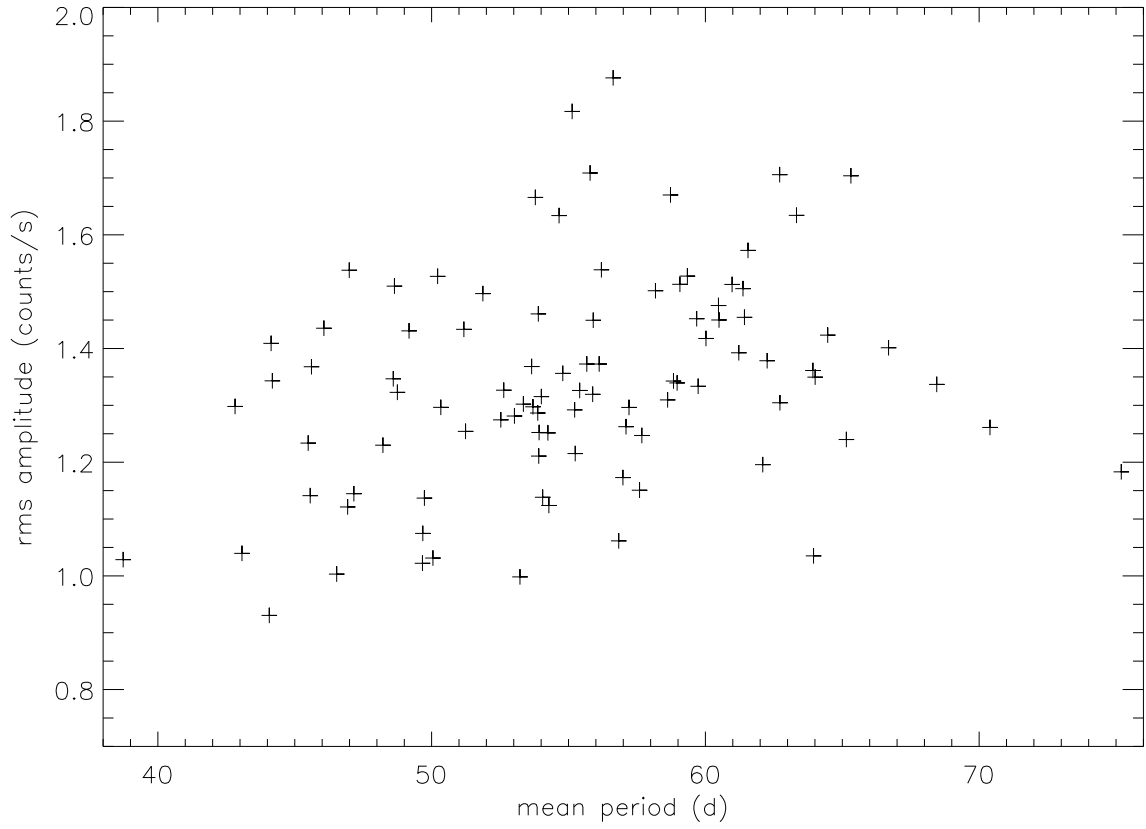}
\caption{Correlation of cycle-averaged period and rms amplitude of original ASM light curve. \label{corr_mp_rmsamp}}
\end{figure}

\end{document}